\documentclass[aps,showpacs,nofootinbib,reprint,pra]{revtex4-1}
\newcommand{\muin}{\mu_\text{in}}
\newcommand{\muout}{\mu_\text{out}}
\newcommand{\muzero}{\mu_\text{0}}
\newcommand{\Pst}{P_\text{st}}
\newcommand{\Psp}{P_\text{sp}}

\newcommand{\Pin}{P_\text{in}}
\newcommand{\Pout}{P_\text{out}}

\usepackage{sistyle}
\usepackage{graphicx}
\usepackage{cancel}

\expandafter\let\csname equation*\endcsname\relax
\expandafter\let\csname endequation*\endcsname\relax\usepackage{amsmath}
\begin{document}
\author{Bruno Sanguinetti}
\email{Bruno.Sanguinetti@unige.ch}
\author{Thiago Guerreiro}
\author{Fernando Monteiro}
\author{Nicolas Gisin}
\author{Hugo Zbinden}
\affiliation{Group of Applied Physics, University of Geneva, Chemin de Pinchat 22, CH-1211 Geneva 4, Switzerland}
\title{Measuring absolute spectral radiance using an Erbium Doped Fibre Amplifier}

\begin{abstract}
We describe a method to measure the spectral radiance of a source in an absolute way without the need of a reference. Here we give the necessary detail to allow for the device to be reproduced from standard fiber-optic components. The device is  suited for fiber-optic applications at telecom wavelengths and calibration of powermeters and spectrometers at light levels from  \SI{1}{nW} to \SI{1}{uW}.
\end{abstract}
\maketitle
\section{Introduction}
Measurement of optical power, although extremely common in both industry and research remains relatively inaccurate, with standard commercial equipment having nominal uncertainties of the order of 5\%.
In a metrology laboratory, an absolute standard such as a cryogenic radiometer can achieve accuracies of \SI{e-4}{} \cite{Martin:2005cz}. This type of experiment however requires considerable time and effort and is usually not available at the site where the measurement is needed. The accuracy of the calibration chain then limits the overall accuracy of the measurement to values which are typically of the order of 1\%. Comparisons of powermeters from different metrology laboratories yield discrepancies of the order of \SI{e-3}{} \cite{Martin:2005cz,Vayshenker:2010vs}.

Furthermore, these accuracies are obtained for relatively high incident powers, of the order of \SI{e-3}{W} which is more than 10 orders of magnitude away form powers used in applications at the quantum level, as a photon at telecom wavelength has an energy of the order of \SI{e-19}{J}.

We have recently demonstrated that the fidelity (degree of polarization at the output) of a cloning process (amplification) in an Erbium doped fibre can be directly related to the amount of input light~\cite{Sanguinetti:2010fya}. This approach, conveniently based on fundamental laws of nature, requires a polarimetric measurement, which when used with commercial equipment limits the precision of the results.

Here we present an alternative method for an absolute measurement which consists in comparing the spontaneous emission of the amplifier $P_\text{sp}$ to the emission stimulated by the source $P_\text{st}$. This method has been implemented using spontaneous parametric downconversion (SPDC) in bulk crystals \cite{Polyakov07,Migdall:1999ut,Polyakov:2009kp} but the free-space nature of the setup makes it challenging to accurately count the number of spatial modes involved. Furthermore the radiances and gain that can be achieved in bulk crystals remain low.

We have implemented this method (on/off technique) and found it to be more practical than the one requiring polarimetry.

\section{Working principle}
\begin{figure*}[htbp]
\begin{center}
\includegraphics[width=1.6\columnwidth]{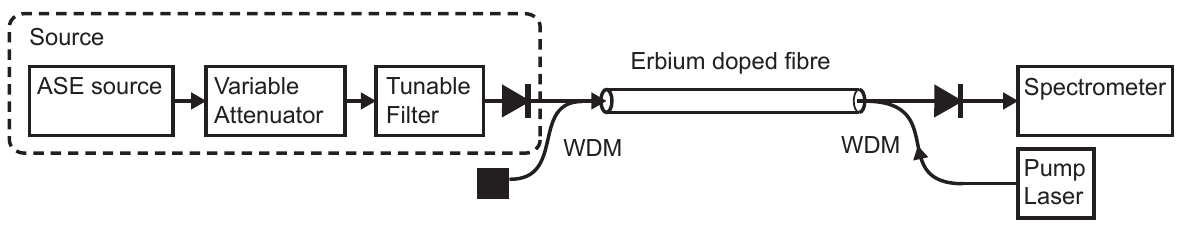}
\caption{Experimental setup: a stable ASE source is filtered and injected into a fully inverted Erbium doped fiber. At the output of this fiber the spectrum is measured with a spectrometer. The power of the source is controlled by a variable attenuator which also contains a shutter. The diode symbols represent optical isolators which both stop back-reflections and remove any remaining \SI{980}{nm} light.}
\label{fig:working_principle}
\end{center}
\end{figure*}
Spontaneous emission in an inverted atomic medium (as well as in parametric processes) can be seen as being stimulated by vacuum fluctuations. These fluctuations are an ``omnipresent standard''~\cite{Migdall:1999ut} and stimulate exactly the same amount of radiation $\muout$ as a radiance $\muin$ of one photon per mode would.

A specific atomic medium, such as a length of Erbium doped fiber, will have a gain $G$ defined in~\cite{Sanguinetti:2010fya} as the increase in output radiance $\muout$ for a given increase in input radiance $\muin$:
\begin{align}
G = \frac{\partial\muout}{\partial\muin}
\label{eq:gain}
\end{align}
The smallest value of gain is $G=1$ representing a medium with no gain, where the output equals the input. Using this definition and assuming no loss, the number of output photons per mode $\muout$
 is:
\begin{align}
\muout = G\,\muin+G-1.
\label{eq:mout_gain}
\end{align}
where the term $G\,\muin$ represent the emission stimulated by the input light and the  term $G-1$ represents the spontaneous emission~$\muzero$.

The radiances above are expressed in number of photons per mode, these unitless values can be converted into an optical power by multiplying them by the photon energy $h\,\nu$ and by the number of modes per second $N$ which is the inverse of the coherence time of the light: $N=1/\tau_c$~\cite{Mandel62}.

Measuring the spontaneous and stimulated emissions with an uncalibrated (but linear) powermeter, the reading will be off by a constant $k$ with respect to the real value so that the measured values are $\Psp^*$ and $\Pst^*$ are respectively:
\begin{align}
\Psp^* &= (G-1)\times \frac{h\nu}{\tau_c}\times k \label{eq:psp}\\
\Pst^* &= (G\,\muin + G - 1) \times \frac{h\nu}{\tau_c}\times k \label{eq:pst}
\end{align}
Eqs. \ref{eq:psp} and \ref{eq:pst} can be solved to obtain $\muin$ independently from the powermeter calibration factor $k$:
\begin{align}
\muin = (1-1/G)(\Pst^*/\Psp^*-1)
\label{eq:muin}
\end{align}
$G$  can be measured exactly with an uncalibrated but linear powermeter so that the measurement of $\muin$ is absolute.
The power $P_\text{in}$ can be evaluated from the measured $\muin$ as:
\begin{align}
\Pin &= \muin \times h\,\nu \times N \\
  &= \muin \times h\,\nu / \tau_c  \\
  &= \muin \times h\,c^2 / \lambda\,l_c
  \label{eq:source_power}
\end{align}
The number of photons per mode $\muin$ is unitless and can therefore be measured absolutely without the need of any other standard. However the measurement of power depends on the coherence time and wavelength of the photons, and is therefore tied to a standard of time or length (Eq.\ref{eq:source_power}), which  can be provided with extreme accuracy by a wavemeter.

We would like to note that although the experimental setup and equipment used in this experiment are different from those of the ``cloning radiometer'' presented in~\cite{Sanguinetti:2010fya} the underlying principles are very similar. In~\cite{Sanguinetti:2010fya} the spontaneous and stimulated emissions were distinguished by their polarization, whereas here they are distinguished in time.

\section{Experimental setup}

In practice an unpolarized, stable source (Amplified Spontaneous Emission) is filtered to define a specific wavelength and bandwidth and then injected into a fully inverted single mode Erbium Doped Fiber where it is amplified.
An uncalibrated but linear powermeter is used to measure the gain of the fiber (a relative measurement) and a spectrometer to measure the relative spectral radiances of the spontaneous and stimulated emission.
The setup is schematically shown in Fig.\ref{fig:working_principle}.

This measurement scheme is particularly robust, as all losses after amplification (including polarization dependent losses) are equal for both the spontaneous and stimulated emission, and therefore cancel as shown above. It is then only necessary to accurately measure the input losses of the device to obtain the spectral radiance of the source in number of photons per mode.
All the fibre used supports a single spatial mode and two polarization modes. Below we will look at each individual element of the radiometer before analyzing the overall setup.

\subsection{Source}
Although this radiometer can operate with a variety of sources such as Light Emitting Diodes or Amplified Spontaneous Emission sources, we have found it useful to develop a reference source which meets the requirements of both the radiometer and subsequent calibration of a powermeter.
The most important aspect of the source is stability, as with a stable source and a stable power meter it is possible to evaluate the stability of all the other elements in the setup. It is also preferable for the source to be unpolarized as the effect of small polarization dependent losses is greatly reduced. Another attractive characteristic for a source is to have a low coherence time so that no interference due to small reflections can occur.

\begin{figure}[h]
\centering
\includegraphics[width=0.7\columnwidth]{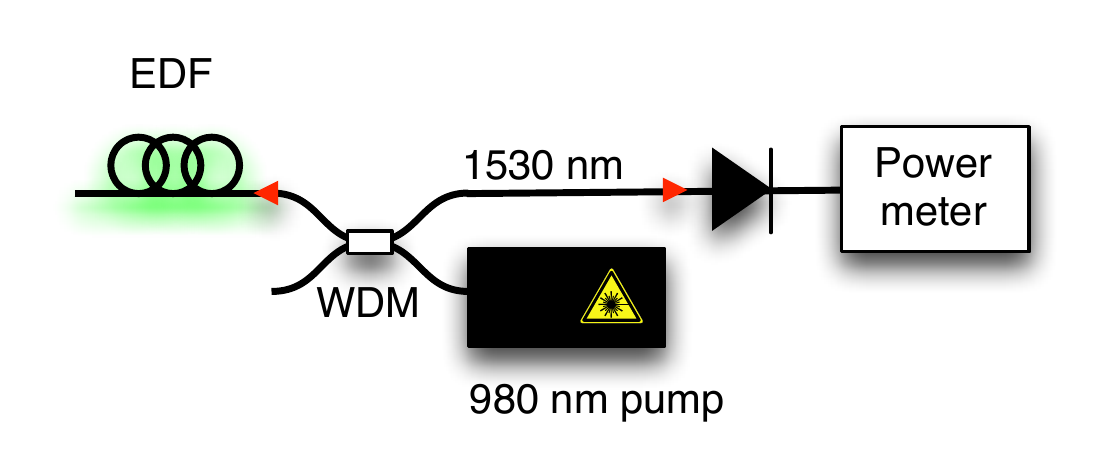}
\caption{Fiber source of spontaneous emission. An Erbium-doped fibre is backwards-pumped, the amplified spontaneous emission counter-propagating to the pump is separated with a wavelength division multiplexer (WDM), any residual pump light is removed using an isolator which has very strong attenuation for the pump light.}
\label{fig:ASE_source_schematic}
\end{figure}

A spontaneous emission source based on a short length of Erbium Doped Fiber (EDF) satisfies these requirements: our setup is shown in Fig.~\ref{fig:ASE_source_schematic}. A \SI{700}{mW} pump laser at \SI{980}{nm} is injected into a wavelength division multiplexer (WDM) and into the Erbium doped fiber.
Most of the pump light then exits the EDF through an angle polished face (back reflection $<\SI{-60}{dB}$). The EDF is kept short in order to generate a sufficiently small amount of light, usable with our radiometer. A short EDF (1mm to 5cm THORLABS ER30-4/125) also ensures that only a small fraction of the pump is absorbed and that the medium is well saturated over its entire length. With such a short fiber, stimulated emission is minimized and the generated light has a degree of polarization $<0.5\%$.

Spontaneous emission at \SI{1530}{nm} passes through the WDM and an isolator which plays the role of a long-pass filter to remove residual pump light. This light is then available for measurement.
This source is very stable as the number of photons produced per second depends on the number of Erbium ions in the fibre and their lifetime, both being constant. Although in practice the medium is only \emph{mostly} inverted , the ASE source is more than three orders of magnitude more stable than the pump laser..

\begin{figure}[h]
\centering
\includegraphics[width=0.95\columnwidth]{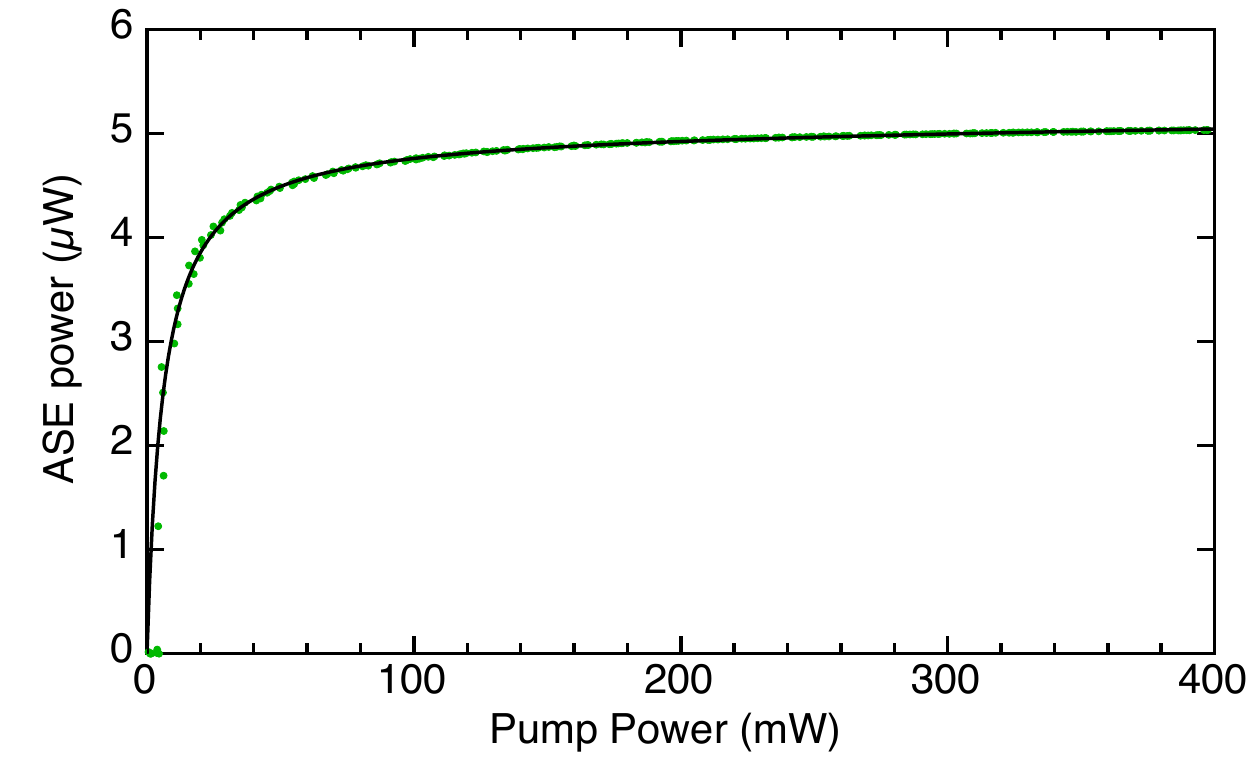}
\label{fig:ASE}
	\caption{
	Saturation of the atomic medium: as the pump laser power $P_\text{p}$ is increased the spontaneous emission power $P_\text{ASE}$ increases as $P_\text{ASE}\propto P_\text{p} / (P_\text{p}+C)$ (fitted curve), the slope $\partial P_\text{ASE} /\partial P_\text{p}$ becomes \SI{2000} times smaller as the pump laser power reaches \SI{400}{mW}.
	}
\end{figure}

When smaller powers are needed, for example to calibrate single photon detectors, a shorter length of erbium-doped fiber can be employed. The amount of amplified spontaneous emission grows exponentially with fiber length, however for short fibers it is approximately linear with length. In order to obtain the small amounts of power required to calibrate single photon detectors it is necessary to use a very short length of Erbium doped fiber ($<\SI{1}{mm})$. This further improves stability, as the medium is more readily saturated. The fiber end is angle polished as to minimize back-reflections of the pump laser.

\begin{figure}[h]
\begin{center}
\includegraphics[width=0.95\columnwidth]{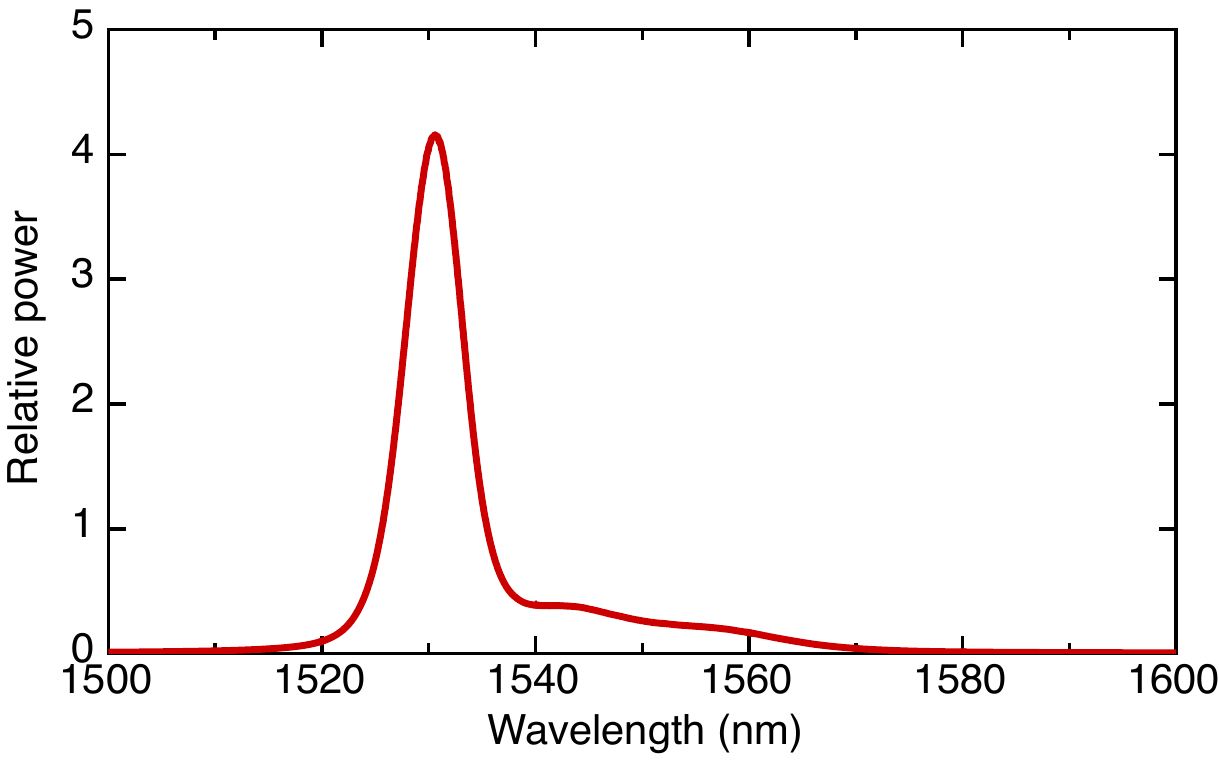}
\caption{Amplified spontaneous emission spectrum, a peak is present at \SI{1530}{nm} and a flat region between \SI{1540}{nm} and \SI{1544}{nm}.}
\label{fig:ASE_spectrum}
\end{center}
\end{figure}

\subsection{Erbium Doped Fiber Amplifier}
The Erbium Doped Fiber Amplifier employed in the radiometer is also of very simple construction. Consisting of a short section of single-mode (\SI{4}{\micro m}) Erbium doped fiber in which the telecom light and a \SI{980}{nm} pump are injected via a WDM. At the output most of the pump light is eliminated using another WDM and an isolator. For our application it is important not to saturate the gain,  especially at the beginning of the amplifier where losses directly influence the measurement. For this reason we chose a gain of approximately 10, a good compromise between low loss and high gain. In principle it is best to pump in the forward direction as to better saturate the fiber at its input, however here we pumped in the backward direction as this allows to verify the absence of any back-reflections as we shall see in the experimental procedure described below.

\subsection{Powermeter}
\begin{figure}[h]
\begin{center}
\includegraphics[width=0.95\columnwidth]{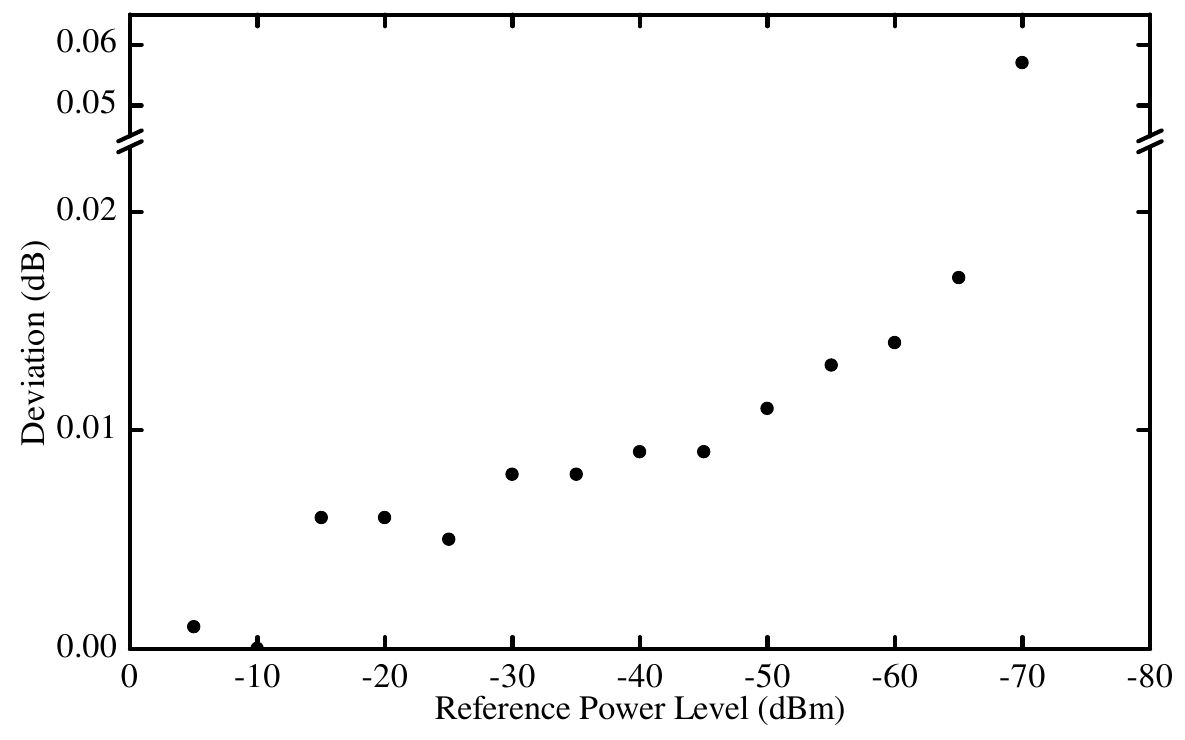}
\caption{Linearity of the powermeter measured by METAS, the Swiss federal office of metrology. The linearity seems good, having a deviation of only 0.4\% between \SI{0}{dBm} and \SI{-65}{dBm}. The last point, at \SI{-70.1}{dBm} has a deviation of 1.3\% mainly due to the background noise of the powermeter.}
\label{fig:powermeter_linearity}
\end{center}
\end{figure}

The radiometer does not require the powermeter  to be calibrated, however it must be stable in time and linear. We used a Thorlabs PM100A with S154C InGaAs fiber head. The stability of the measurement was excellent: when measuring the ASE source, the Allan deviation had a minimum of  \SI{e-5} for timescales of one minute and stayed below \SI{e-4} for days.
The linearity was measured by the Swiss federal office of metrology METAS to have an error of 0.4\% over \SI{60}{dB}, as shown in fig.~\ref{fig:powermeter_linearity}.
%
%
%
The nominal systematic error of this powermeter is 5\%; this relatively high value appears to be given by the presence of reflections and interference between the fiber and the diode.
Custom detectors such as traps~\cite{Fox2005,Fox2000,Lopez2006} would perform much better in this respect.

\subsection{Spectrometer}
\begin{figure}[htbp]
\begin{center}
\includegraphics[width=0.95\columnwidth]{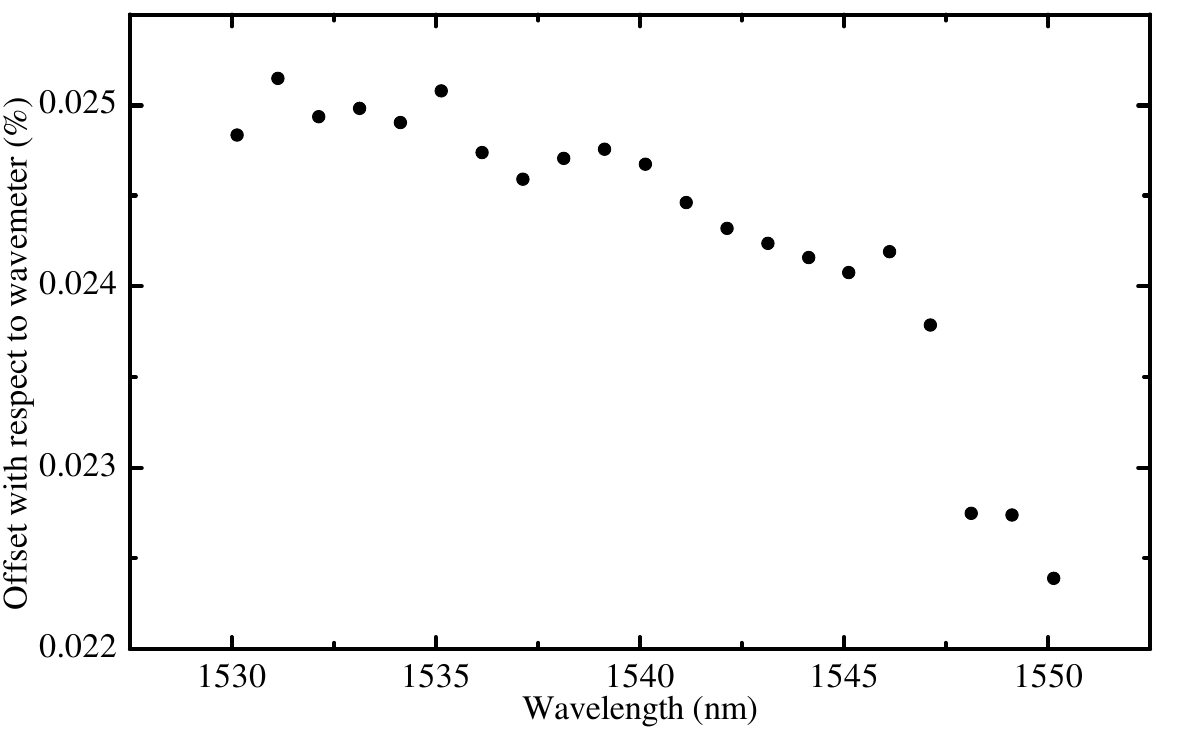}
\caption{Spectrometer calibration curve vs a wavemeter. The calibration error is of the order of \SI{e-4}. Once calibrated (linear fit) the standard deviation of the residuals is \SI{2}{pm}. }
\label{fig:spectrometer_calibration}
\end{center}
\end{figure}

To acquire the spectra we used an Anritsu MS9710C spectrometer which we calibrated against a Bristol Instruments 621 wavemeter. The calibration curve is shown in Fig.~\ref{fig:spectrometer_calibration}. The relative wavelength error due to even an uncalibrated spectrometer is of the order of \SI{e-4}{} and will become relevant only after the powermeter is improved. Once calibrated the error is of the order of \SI{2}{pm}. The linearity of the spectrometer's vertical scale was not tested, however we checked that the offset was zero.

\subsection{Fibers and connectors}
We used standard fibers of two different core diameters at different points of the experiment: where only the telecom (\SI{1542}{nm}) light passes SMF28 fiber was used, whereas in elements which have both telecom and \SI{980}{nm} light we employed fiber with a core diameter of \SI{8}{\micro m} as to guarantee that both wavelengths propagate in a single mode. This results in losses at the interfaces between the two types of fiber\footnote{These interfaces only occur for \SI{1542}{nm} light which propagates in a single mode in both fibers; the overlap between these modes is constant however, i.e. losses are stable.} however the losses of the telecom light due to lesser guiding by the smaller-core diameter fiber were unmeasurable over the short lengths used in the experiment.

In order to achieve best accuracy one would splice all the fibers as connectors (especially APC) can have losses which can change with time due to mechanical and thermal effects. Here however most fibers were connectorized in the interest of flexibility and to better understand the limitations and stabilities of the individual elements of the experiment.

\begin{figure}[h]
\begin{center}
\includegraphics[width=\columnwidth]{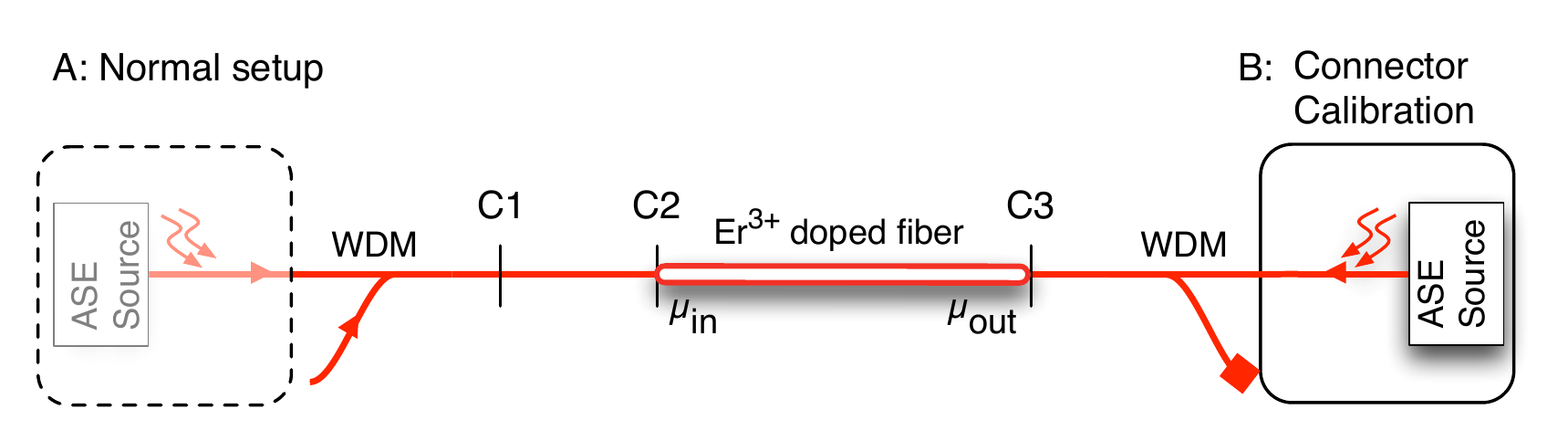}
\caption{In order to calibrate the input losses of the EDF one injects light backwards through connector C3 and measures the amount of light $P_2$ out of the EDF at C2. After this one makes the connection at C2 and measures the amount of light $P_1$ coming out of C1. The input losses are then $P_1/P_2$.}
\label{fig:radiometer_connectors}
\end{center}
\end{figure}

The radiance $\muin$ measured by the radiometer corresponds to amount of light present at the beginning of the EDF. As the radiometer works in a single fiber mode, it is possible to calibrate its input losses by injecting reference light back through the Erbium doped fiber (putting our source in position B of Fig~\ref{fig:radiometer_connectors}) and measuring the losses when making this input connection as illustrated in Fig.~\ref{fig:radiometer_connectors}.

The losses at C2 can be measured precisely, and without introducing random error as this connection is made only once.

\begin{figure}[h]
\begin{center}
\includegraphics[width=0.95\columnwidth]{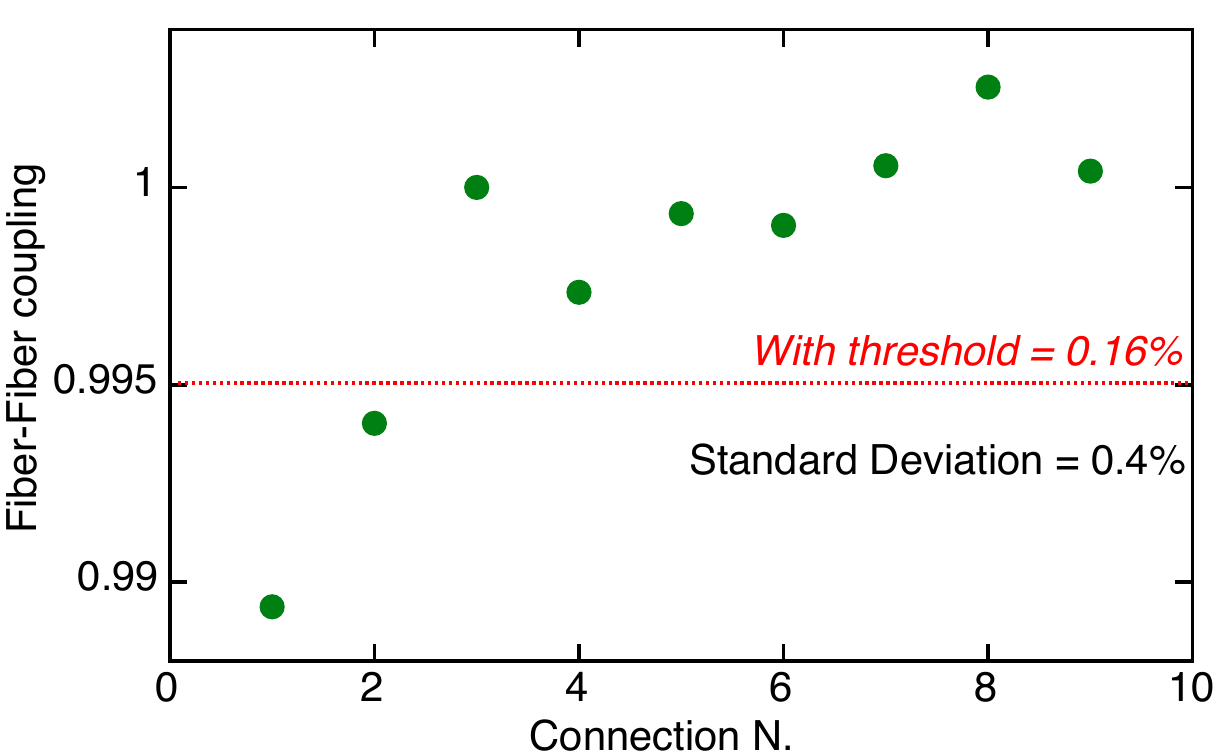}
\caption{Connection statistics. Making a connection without measuring its losses each time will result in a random error. We measured the amplitude of this random error to be 0.4\%. However it can be reduced by only accepting connections with the highest values of transmission. In this case we observe that by rejecting values below a certain threshold we improve the repeatability to 0.16\%. One measurement shows losses slightly higher than 1 (this should not be possible).}
\label{fig:connection_statistics}
\end{center}
\end{figure}

When a radiometric measurement is made, the source is in position A and the reference light with which a powermeter is calibrated is taken from C1. It is therefore important to estimate the losses at this connection. This is done with the source in position A and measuring the amount of light at C1 and at C2. This is the last connection that is made and will introduce a random error. We made this connection multiple times to evaluate its repeatability and found it to be have a standard deviation between consecutive connections of 0.4\%. It is possible to improve on this by only accepting the highest values for transmission as shown in Fig.~\ref{fig:connection_statistics}. Using this method we reduce the error to 0.16\%.

\section{Measurements}
To give a better idea of how a measurement is performed in practice here we go through the experimental procedure in detail.\footnote{The python script where these calculations are performed is attached to this article as supplementary material.}
\subsection{Checking absence of back-reflections} The return loss of fiber connections can be very low if properly done, however it is important to verify that this is below our desired measurement accuracy. For this we disconnect the input of the EDF leaving the APC connector into a beam dump. We turn on the pump laser and measure the amount of light present at the output of the radiometer to be~\SI{10.84}{\micro W} we then connect the input section (C2) and measure the amount of light again to be \SI{10.81}{\micro W}. No back reflections seem to take place in the equipment preceding the C2.
\subsection{Measure input and output losses} To measure output losses light is fed through the EDF with the pump laser turned off. The amount of light at the output of the EDF (C3) is measured to be \SI{14.12}{nW}. Connecting C3 we measure \SI{9.889}{nW} so that the output transmission is evaluated to be $T_\text{out}=\SI{0.7004}{}$.

We evaluate the input losses at connector C1 by measuring the amount of light present in the input fiber before and after making the connection, obtaining \SI{133.3}{nW} and \SI{102.3}{nW} respectively, corresponding to  a transmission of $T_\text{C1}=\SI{0.7674}{}$.The input losses are measured by turning off the pump laser, injecting light from our source backward through the EDF and measuring the amount of light before and after making the C2 connection. We obtain \SI{20.12}{nW} and \SI{17.63}{nW} resulting in $T_\text{C2}=0.8762$. The total input transmission will be $T_\text{in} = T_\text{C1}\times T_\text{C2}=\SI{0.6725}{}$. These relatively high losses are due to the different fiber core diameters employed in the setup.

\subsection{Input power measurement} We measure the input power and spectrum at connector C1 with a standard powermeter for later comparison with the results of the radiometer. Although the powermeter (Thorlabs PM100A with S154C head) is stable to four digits of precision, its reading is influenced by multiple reflections and interference between the fibre connector and detector faces. The reading is also affected by the type of connector used, FC-PC or FC-APC. We evaluated this error to be of the order of 2\% to 4\%.

\begin{figure}[htbp]
\begin{center}
\includegraphics[width=0.95\columnwidth]{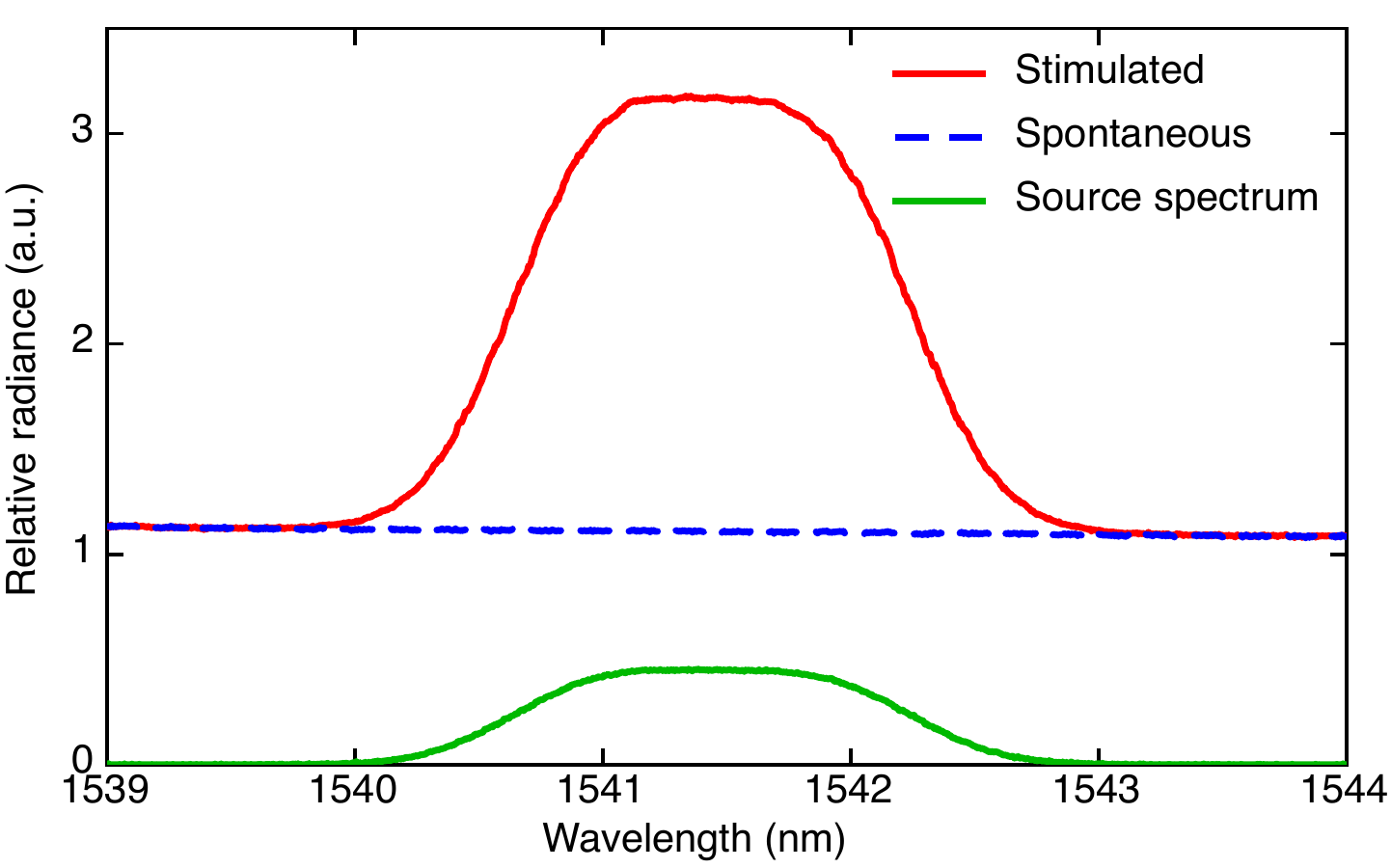}
\caption{Relative spectra of the source light and spontaneous and stimulated emissions}
\label{fig:spectra}
\end{center}
\end{figure}

The spectrum of the light which we inject into the radiometer is shown on fig.~\ref{fig:spectra} and the input power is measured at C1 to be \SI{131.8}{nW}. When calculating the effective amount of light present at the input of the erbium doped fiber, one must take into account the connection losses measured above but also the fact that as the refractive indices of the fibers will match no reflection will occur at the interfaces, this factor represents an increase in transmission of 1.038.

\subsection{Measurement of the gain} The gain is measured by looking at the increase of output power with  input power. In practice we measure the difference in spontaneous and stimulated emission of the radiometer with the powermeter. We obtain \SI{11.29}{\micro W} with the source off and  \SI{11.80}{\micro W}
with the source on. After taking the input and output losses into account we can conclude that the fiber has an average gain of 8.08 over the spectrum of the input light.
In our case the gain is close to flat over the spectrum of the input light, however it is in general possible to deal with non-flat gain, as a spectral measurement of the spontaneous emission of the radiometer can yield the spectral gain as follows. We can rewrite eq.\ref{eq:gain} as:
\begin{align}
G(\lambda) & = \frac{\Delta \Pin(\lambda)}{\Delta \Pout(\lambda)}\\
    & = \muout^\text{sp}(\lambda)+1 \\
    & = k\times y_\text{sp}(\lambda)+1 \label{eq:k}
\end{align}
where $\muout^\text{sp}(\lambda)$ is the radiance of the spontaneous emission of the radiometer and $y_\text{sp}(\lambda)$ is the uncalibrated measurement of this radiance, while $k$ is a calibration constant.
The mean gain $\bar{G}$ over the measured input power spectrum $k\;\Pin(\lambda)$ will be:
\begin{align}
\bar{G} & = \int_0^\infty G(\lambda)\frac{\cancel{k}\Pin(\lambda)}{\cancel{k}\int_0^\infty \Pin(\lambda)d\lambda} d\lambda\\
             & = \int_0^\infty \left[k\, y_\text{sp}(\lambda)+1\right]\frac{\Pin(\lambda)}{\int_0^\infty \Pin(\lambda)\;d\lambda}\;d\lambda
\end{align}
This equation can be solved for $k$:
\begin{align}
k=\frac{\bar{G}-1}{\int_0^\infty\frac{\Pin(\lambda)\,y_\text{sp}(\lambda)}{\int_0^\infty \Pin(\lambda)\;d\lambda}}
\end{align}

The mean gain $\bar{G}$ is measured with the powermeter as described above, whereas  $y_\text{sp}(\lambda)$ and $\Pin(\lambda)$ are measured with the spectrometer. The  $k$ value obtained can then be used in eq.~\ref{eq:k} to get the spectral gain $G(\lambda)$.

\subsection{Calculating the radiance of the source}
Finally, all the parameters calculated above can be put into eq.~\ref{eq:muin} to obtain the radiance. All the raw data files for a typical measurement and the script used to process them are included in the additional information. The measured radiance is shown in fig.~\ref{fig:result}, compared with the radiance measured by the combination of a calibrated powermeter and a spectrometer as described below.

\subsection{Comparison with a calibrated powermeter}

 We compare the spectral radiance $\muin(\lambda)$ measured by the radiometer with that measured by a spectrometer normalized to a calibrated powermeter, calculated as
\begin{align}
\frac{P(\lambda)}{\Delta\lambda} &= \Pin^*(\lambda)\frac{P_\text{tot}}{\int^\infty_{-\infty} \Pin^*(\lambda)\Delta\lambda} \label{eq:power_spectral_density}
\end{align}
where $P_\text{tot}$ is measured with a powermeter and serves to normalize the powers $\Pin^*(\lambda)$ measured for each spectrometer `pixel' of width $\Delta\lambda$.

The radiance is written in number of photons per mode as $\mu=J/N$ where $J$ is the number of photons and $N$ the number of modes in a  bandwidth $d\nu$ and per unit time. The number of photons can be estimated as the measured power $P(\lambda)$ in the frequency range $\Delta\nu$ around $\lambda$ (corresponding to a `pixel' of the spectrometer) divided by the photon energy $h\nu$, so that $J=P(\lambda)/h\nu$. The number of temporal modes per second is the inverse of the coherence time $1/\tau_c=\Delta\nu$ \cite{Mandel62}.
The fibers we use support a single spatial mode but two polarization modes so that the total number of modes per second is $N=2\times\Delta\nu$. The number of photons per temporal mode in a small frequency bandwidth $\Delta\nu$ around frequency $\nu$ can then be calculated from the measured power spectral density $P(\lambda)/\Delta\nu$:
\begin{align}
\mu(\lambda) &=\frac{P(\lambda)}{2\,h\nu\,\Delta\nu}\\
       &=\frac{P(\lambda)\lambda^3}{2\,hc^2\,\Delta\lambda} \ \label{eq:radiance_from_coherence_time}
\end{align}
It is interesting to note that the same result can be obtained with a different approach, detailed in Appendix A, which also includes a strict derivation of why the number of modes per second is $\Delta\nu$.

\begin{figure}[htbp]
\begin{center}
\includegraphics[width=0.95\columnwidth]{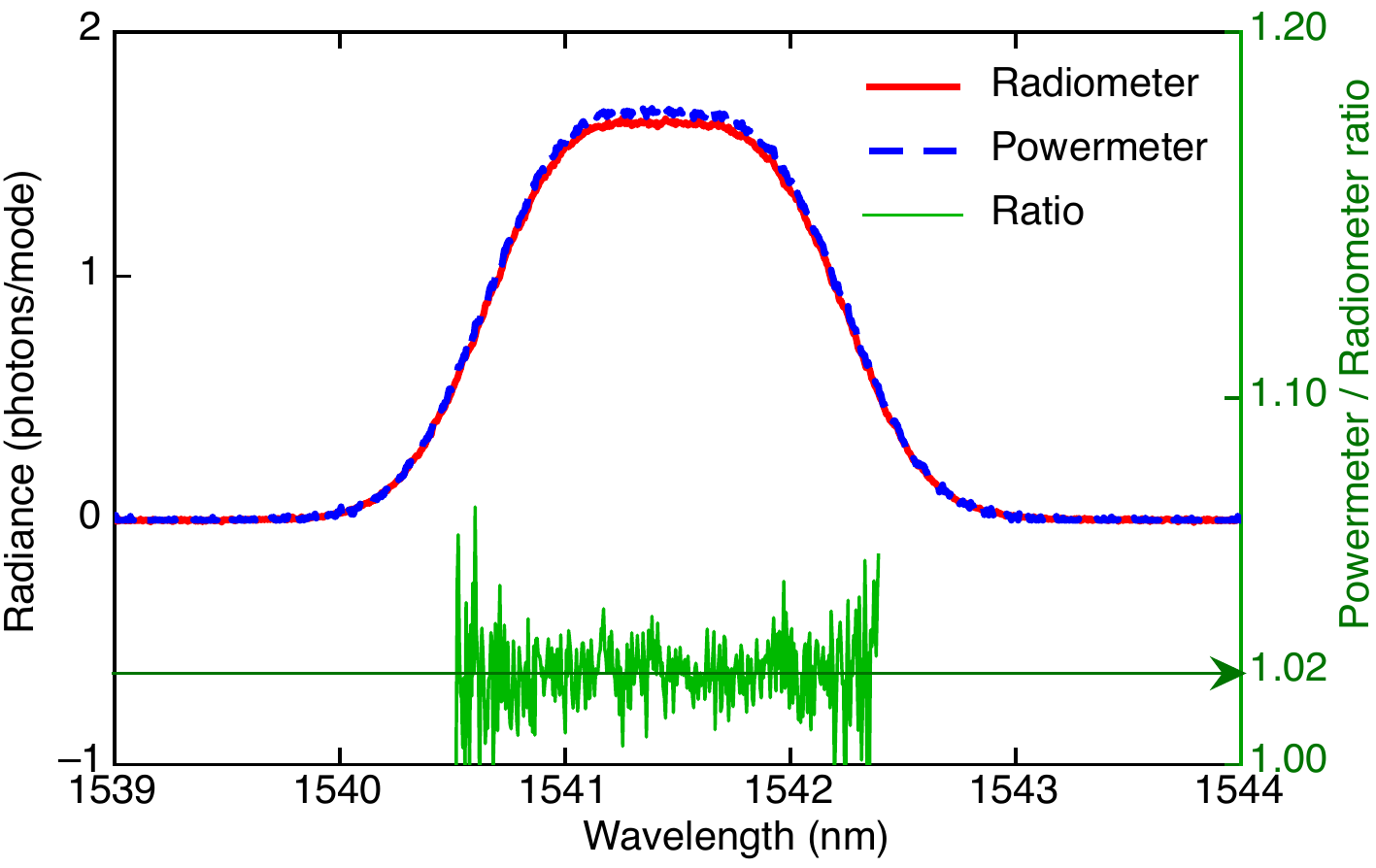}
\caption{Comparison between the spectral radiance measured with the powermeter and with the radiometer.}
\label{fig:result}
\end{center}
\end{figure}

As described above, $P(\lambda)$ is calculated by multiplying the power measured by the powermeter by the normalized spectrum acquired with the spectrometer. The results are shown on fig.~\ref{fig:result}. The radiometer measures a radiance which is 2\% lower than that measured by the spectrometer/powermeter.

\section{Discussion}
Although the absolute radiance measured with the radiometer agrees with that measured with a powermeter/spectrometer combination (which have a 5\% nominal error), to rule out systematic errors, this kind of measurement should be done with the techniques and equipment of a metrology laboratory. We believe that the comparison with national standards will be interesting but will leave this as the subject of future work.
Here however we can discuss the results in terms of linearity and stability, which we have measured using a slightly different technique: instead of the spectrometer discussed above we selected a narrow spectral bandwidth using a filter, the rest of the treatment remaining identical.

\begin{figure}[h]
\begin{center}
\includegraphics[width=0.95\columnwidth]{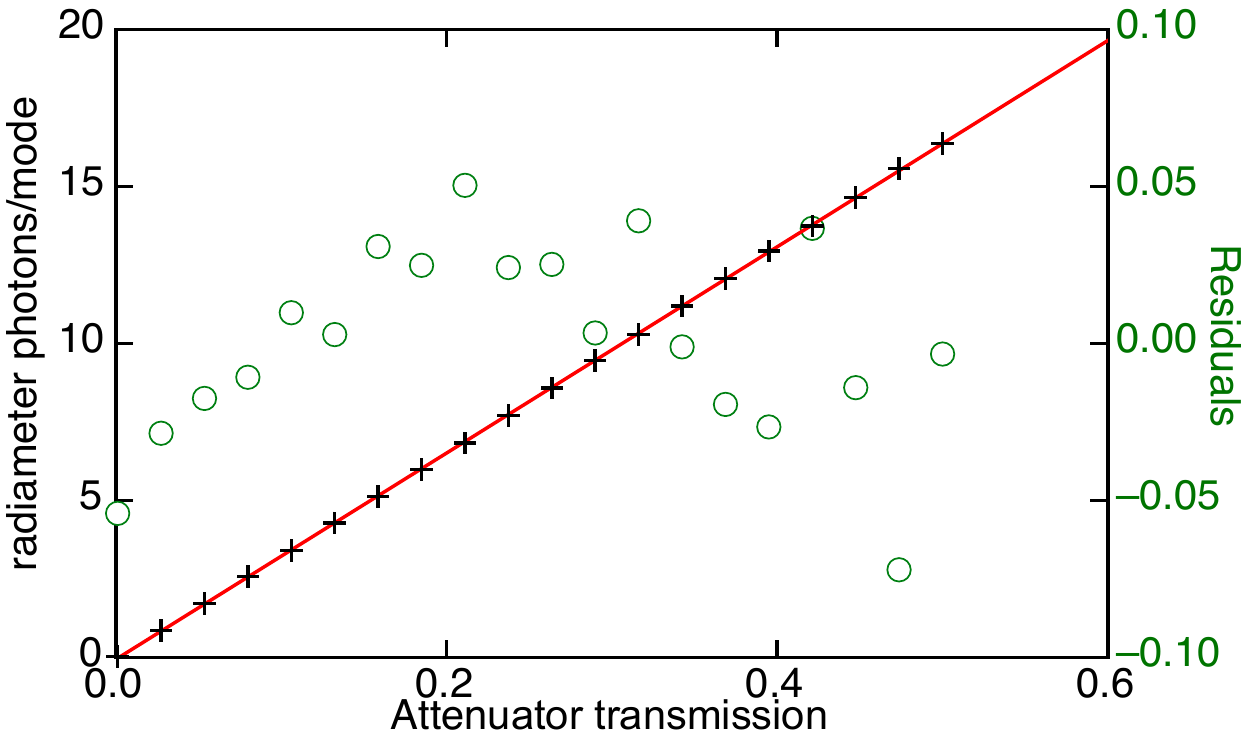}
\caption{Linearity of our radiometer. Number of photons per mode measured by our radiometer versus transmission of the variable attenuator (linear scale). Residuals are magnified 100x and show no sign of saturation.}
\label{fig:linearity}
\end{center}
\end{figure}

\begin{figure}[h]
\begin{center}
\includegraphics[width=0.95\columnwidth]{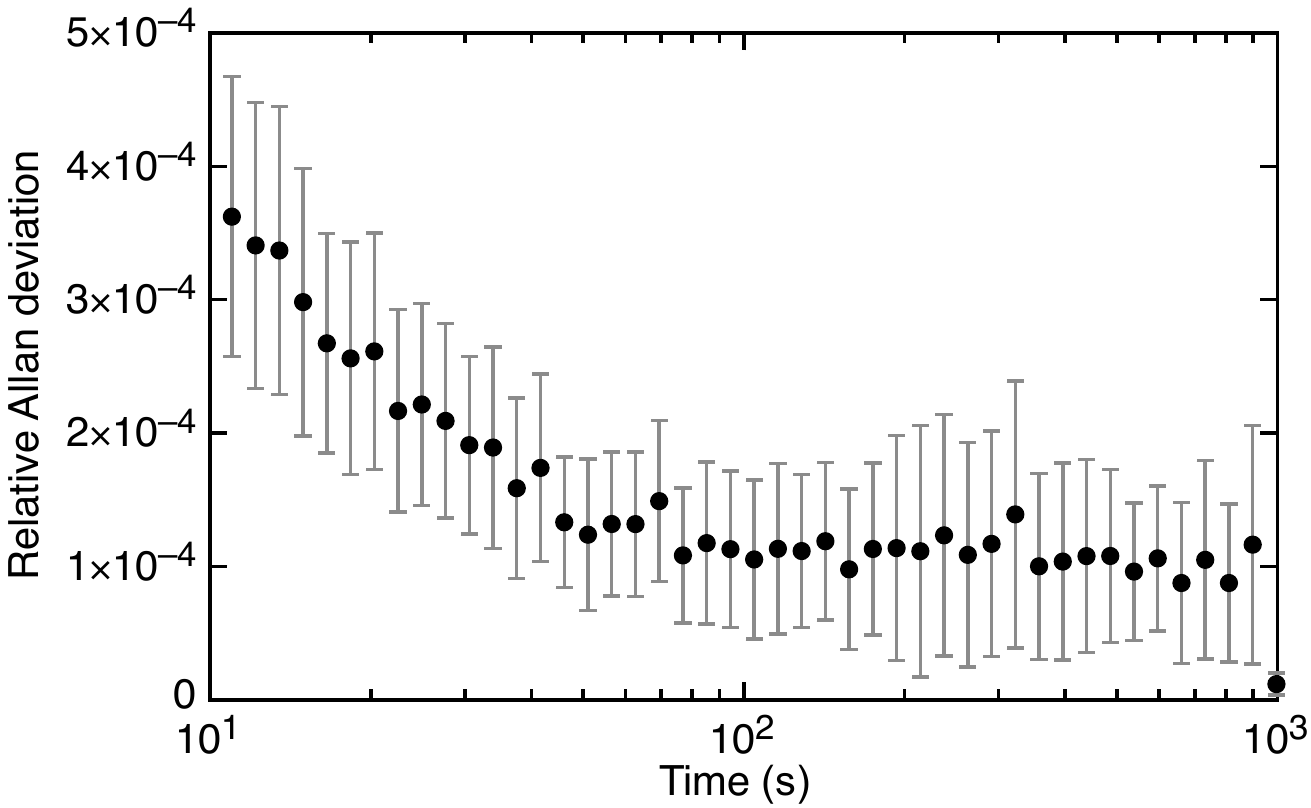}
\caption{Stability over a period of two hours: the Allan deviation drops to a value of \SI{e-4}.}
\label{fig:allan_deviation}
\end{center}
\end{figure}

\subsection{Gain saturation and linearity}
This method, unlike the radiometer described in~\cite{Sanguinetti:2010fya}, requires the entire erbium doped fiber to be fully inverted under all operating conditions. Excess input light could deplete the number of atoms in the exited state and result in losses that are only present when $\Pst$ is being measured, and not when measuring $\Psp$. This nonlinearity will result in a systematic error to our measurement, and although it can be accounted for, it is best avoided. Figure~\ref{fig:linearity} shows that our device is linear up to an input radiance of 15 photons/mode (the optimal operating point for the radiometer is 1 photon/mode, or $\approx$\SI{7}{nW}).
The stability of the radiometer was measured over two hours and found to be of \SI{e-4} as can be seen in figure \ref{fig:allan_deviation}.

\subsection{Error estimation}
Evaluating systematic errors is difficult. In particular some sources of error may remain unnoticed. Moreover loss mechanisms which we fail to take into account will influence our measurement in the same direction, i.e. we will always underestimate the input radiance.

Table \ref{table:uncertainties} includes an estimate of the error sources which we have evaluated which show that this system has the potential of making an absolute measurement of radiance with an accuracy better than 1\%. A better error analysis requires the resources of a metrology laboratory. We  hope  that this paper stimulates further work by a specialized team.

\begin{table}
\centering
\begin{tabular}{lllll}
\multicolumn{1}{r}{Parameter} & \multicolumn{1}{c}{Value} & \multicolumn{1}{c}{Uncertainty} & \multicolumn{1}{c}{Scaling} & \multicolumn{1}{c}{Effect} \\
\hline
\multicolumn{1}{r}{$G$} & \multicolumn{1}{c}{8.08} & \multicolumn{1}{c}{2\%} & \multicolumn{1}{c}{1/G} & \multicolumn{1}{c}{0.25\%} \\
\multicolumn{1}{r}{$y_\text{st}/y_\text{sp}$} & \multicolumn{1}{c}{2.9} & \multicolumn{1}{c}{$\approx\SI{e-4}{\%/dB}$} & \multicolumn{1}{c}{$y_\text{st}/y_\text{sp}$} & \multicolumn{1}{c}{\SI{40}{ppm}} \\
\multicolumn{1}{r}{Insertion loss} & \multicolumn{1}{c}{0.67} & \multicolumn{1}{c}{0.16\%} & \multicolumn{1}{c}{Linear} & \multicolumn{1}{c}{0.16\%} \\
\multicolumn{1}{r}{Wavelength*} & \multicolumn{1}{c}{\SI{1541}{nm}} & \multicolumn{1}{c}{\SI{2}{pm}} & \multicolumn{1}{c}{$\lambda^3$} & \multicolumn{1}{c}{\SI{4}{ppm}} \\
\multicolumn{1}{r}{$\Delta\lambda$*} & \multicolumn{1}{c}{\SI{5}{pm}} & \multicolumn{1}{c}{\SI{100}{ppm}} & \multicolumn{1}{c}{Linear} & \multicolumn{1}{c}{\SI{100}{ppm}} \\
\multicolumn{1}{r}{PM Calibration*} & \multicolumn{1}{c}{0.98} & \multicolumn{1}{c}{5\%} & \multicolumn{1}{c}{Linear} & \multicolumn{1}{c}{5\%} \\
\end{tabular}
\caption{Summary of the uncertainties which play a direct role into our measurement of radiance. The starred (*) items only play a role when comparing our measured value of radiance with that of a powermeter.}
\label{table:uncertainties}
\end{table}

\section{Conclusion}
In this work we have shown how a device capable of measuring absolute spectral radiance of a source can be built simply from standard telecom equipment. We discussed the details of the implementation and shown how limiting factors such as gain saturation and input losses estimation can be dealt with. We believe that this work provides sufficient detail to be reproduced independently and in particular to be developed and validated by a metrology laboratory.

\section{Acknowledgments}
We are very grateful to Jacques Morel and Armin Gambon from the Swiss federal office for metrology (METAS) for useful discussion, for hints and ideas and for calibrating our powermeter. We are also grateful to Silke Peters and Stefan K\"{u}ck from the Physikalisch-Technische Bundesanstalt (PTB), Christopher Chunnilall from NPL and Alan Migdall from NIST for useful discussion.
We thank the Swiss NCCR QSIT for financial support.
\appendix
\section{Radiance measured with a powermeter}
In the text we estimate the radiance of our source using a powermeter. The calculation relies on the definition of coherence time to count the number of modes per second. It is interesting to note that the same result can be obtained with a different approach. Consider Planck's blackbody radiation
\begin{align}
B_\lambda(T)&=\frac{2\,h\,c^2}{\lambda^5}\frac{1}{e^{hc/\lambda\,k\,T}-1}
\end{align}
The last term can be identified as the number of photons per mode. Hence, the spectral radiance (in units of $\SI{}{W/ ( sr .m^2 .Hz)}$) of a source with $\mu$ photons per mode can be written as
\begin{align}
L_\lambda(\mu)=\frac{2\,h\,c^2}{\lambda^5}\mu
\end{align}
To obtain the spectral power density we multiply $L_\lambda$ by the beam surface $\pi\,\omega^2$ and the solid angle $\pi\,\theta^2$ which for a Gaussian beam is $\lambda^2$.
We can write this power spectral density as
\begin{align}
\frac{P(\lambda)}{\Delta\lambda} &= \frac{2\,h\,c^2}{\lambda^3}\mu \\
\text{where:}
P(\lambda) &= P_\text{tot}\frac{\Pin^*(\lambda)d\lambda}{\int^\infty_{-\infty} \Pin^*(\lambda)d\lambda}
\end{align}
So that
\begin{align}
P_\text{tot} &= \mu(\lambda)\frac{2\,h\,c^2}{\lambda^3}\frac{\int^\infty_{-\infty} \Pin^*(\lambda)d\lambda}{\Pin^*(\lambda)}
\end{align}
Which using eq.~\ref{eq:power_spectral_density} can be rewritten as:
\begin{align}
\mu(\lambda)&=\frac{P(\lambda)\lambda^3}{2\,hc^2\,\Delta\lambda},
\end{align}
identical to eq.~\ref{eq:radiance_from_coherence_time}.

\section{Coherence time of a frequency element $\Delta\nu$}
\begin{figure}
  \centering
  \includegraphics[width=0.7\columnwidth]{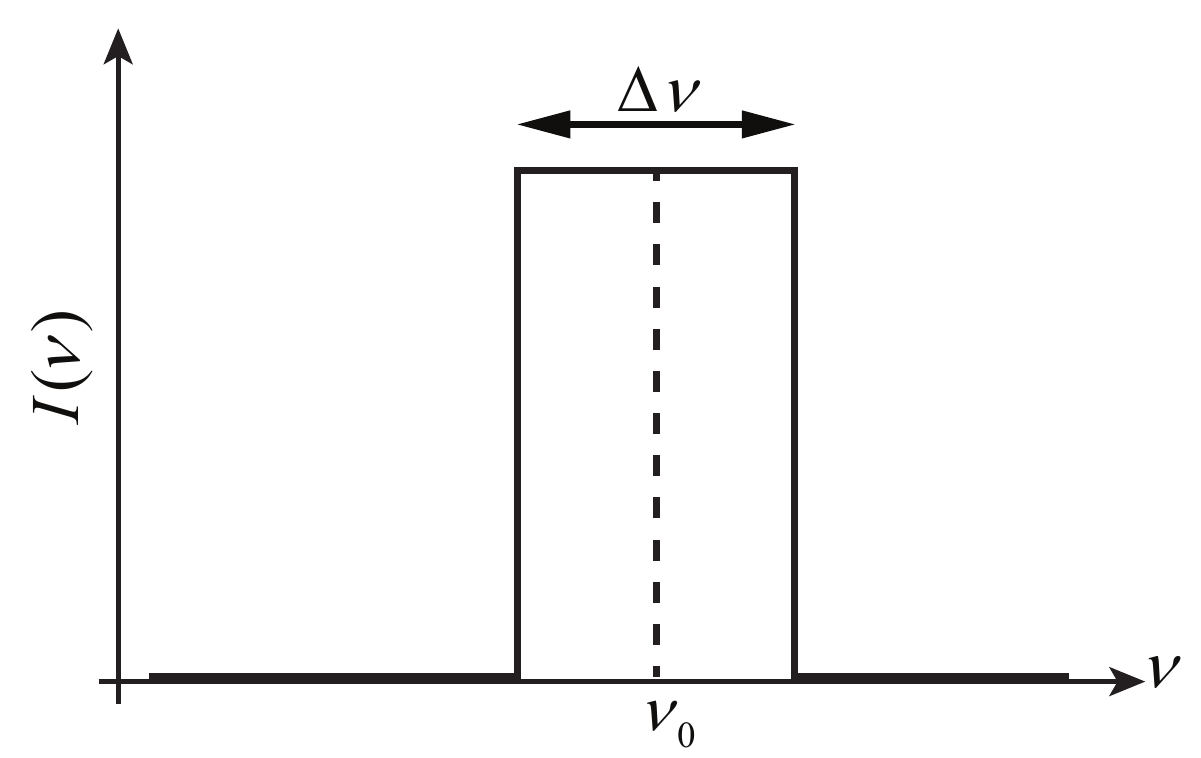}\\
  \caption{Power spectral density function $I(\nu)$ for a small frequency element $\Delta\nu$ centered at $\nu_0$. }\label{fig:square_power_spectral_density}
\end{figure}

Although we can trivially state that a frequency element $\Delta\nu$ has a coherence time of $1/\Delta\nu$, it is reassuring to derive this using a more general approach.
Fig.~\ref{fig:square_power_spectral_density} shows the power spectral density of a small frequency element $\Delta\nu$ of our signal. We define this element as having a square shape which guarantees that there is no overlap between adjacent elements. The coherence time $\tau_0$ is define in terms of the autocorrelation function $\gamma(t)$ of the signal such that
\begin{equation}\label{eq:autocorrelation_function_definition}
  \tau_0 = \int^\infty_{-\infty}|\gamma(t)|^2 dt.
\end{equation}
This definition of $\tau_c$ in not arbitrary but is a measurable physical quantity tied to the length $c\,\tau_c$ of the unit cell of photon phase-space as described in~\cite{Mandel62}.
The autocorrelation function $\gamma(t)$ is the fourier transform of the power spectral density $I(\nu)$ so that:
\begin{align}\label{eq:autocorrelation_derivation}
\gamma(t) &= \int^\infty_{-\infty}I(\nu)e^{2\pi i t \nu}d\nu \\
          &= \frac{\sin(\pi t \Delta\nu)}{\pi t \Delta\nu}
\end{align}
Integrating this function we obtain the coherence time:
\begin{align}
\tau_0 &= \int^\infty_{-\infty}\left|\frac{\sin(\pi t \Delta\nu)}{\pi t \Delta\nu}\right|^2 dt\\
       &= \frac{1}{\Delta\nu}
\end{align}
which is the trivial result that we expected.

\bibliographystyle{unsrt}	
\bibliography{Radiometer_bibliography_2}

\end{document}